\documentstyle[12pt]{article}
\def\th{\theta}

\def\ka{\kappa}

\def\si{\sigma}

\def\ph{\phi}

\def\ps{\psi}

\def\Ps{\Psi}

\def\fr#1#2{{{#1} \over {#2}}}

\def\vev#1{\langle {#1}\rangle}

\def\frac#1#2{{\textstyle{{#1}\over {#2}}}}

\def\lsim{\mathrel{\rlap{\lower4pt\hbox{\hskip1pt$\sim$}}
    \raise1pt\hbox{$<$}}}
\def\gsim{\mathrel{\rlap{\lower4pt\hbox{\hskip1pt$\sim$}}
    \raise1pt\hbox{$>$}}}
\def\sqr#1#2{{\vcenter{\vbox{\hrule height.#2pt
         \hbox{\vrule width.#2pt height#1pt \kern#1pt
         \vrule width.#2pt}
         \hrule height.#2pt}}}}

\newcommand{\beq}{\begin{equation}}
\newcommand{\eeq}{\end{equation}}
\newcommand{\bea}{\begin{eqnarray}}
\newcommand{\eea}{\end{eqnarray}}
\newcommand{\rf}[1]{(\ref{#1})}

\renewenvironment{thebibliography}[1]
 { \rm
   \begin{list}{\arabic{enumi}.}
    {\usecounter{enumi} \setlength{\parsep}{0pt}
     \setlength{\itemsep}{3pt} \settowidth{\labelwidth}{#1.}
     \sloppy
    }}{\end{list}}

\begin{document}
\titlepage

\begin{flushright}
{COLBY-96-04\\}
{IUHET 338\\}
{July 1996\\}
\end{flushright}
\vglue 1cm
	    
\begin{center}
{{\bf WAVE-PACKET REVIVALS FOR QUANTUM SYSTEMS\\
WITH NONDEGENERATE ENERGIES\\}
\vglue 1cm
{Robert Bluhm,$^a$ V. Alan Kosteleck\'y,$^b$ and
Bogdan Tudose$^b$\\} 
\bigskip
{\it $^a$Physics Department\\}
\medskip
{\it Colby College\\}
\medskip
{\it Waterville, ME 04901, U.S.A.\\}
\bigskip
{\it $^b$Physics Department\\}
\medskip
{\it Indiana University\\}
\medskip
{\it Bloomington, IN 47405, U.S.A.\\}

}
\vglue 0.8cm

\end{center}

{\rightskip=3pc\leftskip=3pc\noindent
The revival structure of wave packets is examined 
for quantum systems having energies that depend on
two nondegenerate quantum numbers.
For such systems,
the evolution of the wave packet is controlled by
two classical periods and three revival times.
These wave packets exhibit quantum beats in the
initial motion as well as new types of long-term revivals.
The issue of whether fractional revivals can
form is addressed.
We present an analytical proof showing that at
certain times equal to rational fractions of the
revival times the wave packet can reform as a sum
of subsidiary waves and that both conventional 
and new types of fractional revivals can occur.

}

\vskip 1truein
\centerline{\it Accepted for publication in Physics Letters A}

\vfill
\newpage

\baselineskip=20pt

Among the distinctive quantum features of 
wave-packet evolution in a potential
is the formation of fractional revivals.
These appear when the packet reforms 
as a sum of distinct subsidiary waves
some time after its initial collapse.
Fractional revivals have been observed experimentally
in the motion of atomic Rydberg wave packets
\cite{yeazell3,meacher,wals} 
and in molecular vibrational wave packets
\cite{stolow}.
In these systems,
the wave packet is formed via 
excitation with a short laser pulse,
and its motion is detected using a pump-probe configuration
\cite{az}.
The revival structure of a free hydrogenic wave packet 
is determined by the dependence 
of the component-eigenstate energies 
on the principal quantum number.
Similarly,
the vibrational states of a molecular wave packet 
depend only on the vibrational quantum number.

For the case of quantum systems with energies 
depending only on one quantum number $n$,
a generic proof has been given
showing that fractional revivals form
\cite{ap}.
The proof applies to packets that are a 
superposition of states strongly peaked 
around a central value $\bar n$ of $n$.
Initially,
the motion of a wave packet of this type is periodic with 
period $T_{\rm cl}$ equal to the period of the
corresponding classical motion.
After several classical periods,
the wave packet spreads and collapses.
The subsequent formation of revivals is
characterized by a time scale $t_{\rm rev}$
called the revival time.
Fractional revivals occur at times that are 
rational fractions of $t_{\rm rev}$.
The associated subsidiary wave packets 
exhibit a periodicity in their motion
equal to a fraction of $T_{\rm cl}$.

A full revival can be viewed as a special fractional revival 
occurring when the wave packet reforms as a single entity 
evolving with the classical periodicity.
For quantum systems with energies that are quadratic in $n$,
the full revivals are perfect in the sense that 
the wave function at time $t = t_{\rm rev}$ 
is exactly equal to the wave function at $t = 0$
\cite{bkp}.
For other systems the revivals at $t = t_{\rm rev}$ are imperfect,
and new structure emerges after several cycles 
of collapse and revival. 
A detailed analysis involves introducing 
a superrevival time scale $t_{\rm sr}$ and
showing that the interplay between $t_{\rm sr}$
and $t_{\rm rev}$ results in a new sequence of
fractional and full superrevivals for $t \gg t_{\rm rev}$
\cite{sr}.
Among these,
on time scales much larger than
the superrevival time scale $t_{\rm sr}$,
are the long-term revivals 
already predicted in early work on Rydberg wave packets
\cite{ps}
that appear when the phase in the wave function 
is an integer multiple of $2 \pi$
\cite{peres}.

The above analyses can be applied provided the physics
of the wave packet is controlled by a single quantum number.
As an example, 
consider a radial Rydberg wave packet in an alkali-metal atom.
The energies of the atom have quantum defects
and hence depend on more than one quantum number.
However, 
for large enough principal quantum number and fixed 
angular-momentum quantum number,
the quantum defects are constant to an excellent approximation.
The fractional revival and superrevival structure
of a wave packet formed as a superposition of, 
for example, 
p states can therefore be treated using the above methods
\cite{sr}
taking the quantum-defect shifts into account analytically
\cite{sqdt}.
A theoretical description of these wave packets
as squeezed states can be used to extend the treatment
to elliptical wave packets in alkali-metal atoms 
\cite{squeezed}.

To study more complicated cases,
such as Rydberg wave packets in external fields,
an analysis is required that allows for 
nondegenerate energies 
depending on more than one quantum number.
This is the topic of the present paper.

We focus attention here on the dynamics of a wave packet 
$\Ps (t)$ formed as a coherent superposition
of states $\ph_{n_1 n_2}$ with energies $E_{n_1 n_2}$
depending on two quantum numbers $n_1$ and $n_2$:
\beq
\Psi (t) = \sum_{n_1, n_2} c_{n_1 n_2}
\ph_{n_1 n_2} \exp \left[ -i E_{n_1 n_2} t \right]
\quad .
\label{psit}
\eeq
Here, 
$c_{n_1 n_2}$ are complex weighting coefficients.
Note that $n_1$ and $n_2$ may represent effective
quantum numbers in,
for instance,
a perturbative treatment,
for which the states $\ph_{n_1 n_2}$ 
may be only approximate energy eigenstates.
Our analysis in the present work
is therefore relevant for experiments
studying the behavior of Rydberg wave packets in
external fields,
where the energy degeneracy 
in the superposition of states is lifted
\cite{noordam,walther,broers,lankhui,bucksbaum}.

We suppose that 
the superposition is weighted around central values
$\bar n_1$ and $\bar n_2$ of the two quantum numbers.
The weighting coefficients $c_{n_1 n_2}$
are therefore taken to be strongly peaked 
around a central value of the
energy $E_{\bar n_1 \bar n_2}$,
as would result from an excitation process 
using short laser pulses.
The energy can then be expanded in a Taylor series
as
\bea
E_{n_1 n_2} \simeq E_{\bar n_1 \bar n_2} \,
+ \, \left( \fr {\partial E} {\partial n_1} 
\right)_{\bar n_1,\bar n_2} (n_1 - \bar n_1) \,\,
+ \, \left( \fr {\partial E} {\partial n_2} 
\right)_{\bar n_1,\bar n_2} (n_2 - \bar n_2)
\cr
\quad\quad\quad\quad
+ \fr 1 2 \left( \fr {\partial^2 E} {\partial n_1^2} 
\right)_{\bar n_1,\bar n_2} (n_1 - \bar n_1)^2 \,\,
+ \, \fr 1 2 \left( \fr {\partial^2 E} {\partial n_2^2} 
\right)_{\bar n_1,\bar n_2} (n_2 - \bar n_2)^2
\cr
\quad\quad\quad\quad\quad
+ \left( \fr {\partial^2 E} {\partial n_1 \partial n_2} 
\right)_{\bar n_1,\bar n_2} (n_1 - \bar n_1) (n_2 - \bar n_2)
+ \ldots
\quad .
\label{Taylor}
\eea
Since the energy depends on two variables,
the expansion contains a mixed derivative term
at second order.

When Eq.\ \rf{Taylor} is substituted into \rf{psit}
and an irrelevant overall phase is disregarded,
it follows that each derivative term corresponds to 
a distinct time scale.
We define
\beq
T_{\rm cl}^{(1)} 
= \fr {2 \pi} {\left( \fr {\partial E} {\partial n_1} 
\right)_{\bar n_1,\bar n_2}}
\quad , \quad\quad\quad
T_{\rm cl}^{(2)} 
= \fr {2 \pi} {\left( \fr {\partial E} {\partial n_2} 
\right)_{\bar n_1,\bar n_2}}
\quad ,
\label{Tcldef}
\eeq
\beq
t_{\rm rev}^{(1)} = \fr {2 \pi} {\fr 1 2 \left( 
\fr {\partial^2 E} {\partial n_1^2} 
\right)_{\bar n_1,\bar n_2}}
\quad , \quad\quad\quad
t_{\rm rev}^{(2)} = \fr {2 \pi} {\fr 1 2 \left( 
\fr {\partial^2 E} {\partial n_2^2} 
\right)_{\bar n_1,\bar n_2}}
\quad ,
\label{trevdef}
\eeq
\beq
t_{\rm rev}^{(1 2)} = \fr {2 \pi} {\left( 
\fr {\partial^2 E} {\partial n_1 \partial n_2} 
\right)_{\bar n_1,\bar n_2}}
\quad ,
\label{trev12def}
\eeq
where we have assumed for simplicity that the
quantum numbers take integer-spaced values.
For each quantum number,
there is a classical period and a revival time scale.
In addition,
the mixed derivative term generates
a third time scale $t_{\rm rev}^{(1 2)}$,
which we call the cross-revival time.
This term has no analog in the conventional revival
structure of free Rydberg wave packets.
Keeping terms to second order then gives 
for $\Psi(t)$ the expression
\beq
\Psi (t) = \sum_{k_1, k_2} c_{k_1 k_2}
\ph_{k_1 k_2} \exp \left[ -2 \pi i
\left(  \fr {k_1 t} {T_{\rm cl}^{(1)}}
+ \fr {k_2 t} {T_{\rm cl}^{(2)}}
+ \fr {k_1^2 t} {t_{\rm rev}^{(1)}}
+ \fr {k_2^2 t} {t_{\rm rev}^{(2)}}
+ \fr {k_1 k_2 t} {t_{\rm rev}^{(12)}}
\right)
\right]
\quad ,
\label{psiexpans}
\eeq
where $k_1 = (n_1 - \bar n_1)$, $k_2 = (n_2 - \bar n_2)$,
and where the coefficients $c_{k_1 k_2}$ and eigenstates
$\ph_{k_1 k_2}$ are related to the analogous quantities 
in Eq.\ \rf{psit} by shifts in the subscripts.

Consider the time evolution of the wave packet.
When $t$ is small,
the dominant terms in the time-dependent phase in 
Eq.\ \rf{psiexpans} are the first two.
They produce beating between the classical periods
$T_{\rm cl}^{(1)}$ and $T_{\rm cl}^{(2)}$.
The latter are said to be commensurate if
\beq
T_{\rm cl}^{(1)} = \fr a b T_{\rm cl}^{(2)}
\quad ,
\label{relab}
\eeq
where $a$ and $b$ are relatively prime integers.
When the classical periods are commensurate,
the wave-packet displays a period on short time scales
given by 
\beq
T_{\rm cl} = b T_{\rm cl}^{(1)} = a T_{\rm cl}^{(2)}
\quad .
\label{Tcl}
\eeq
If two classical periods are incommensurate,
the initital motion of $\Ps (t)$ is not exactly periodic.

On larger time scales
the three quantities 
$t_{\rm rev}^{(1)}$, $t_{\rm rev}^{(2)}$, and
$t_{\rm rev}^{(12)}$ 
play a role in controlling the behavior,
eventually inducing the wave-packet collapse.
It follows from the form of 
the three second-order terms in the time-dependent phase
that full revivals appear
if the three revival times 
$t_{\rm rev}^{(1)}$, $t_{\rm rev}^{(2)}$, and
$t_{\rm rev}^{(12)}$
are commensurate and satisfy 
\beq
t_{\rm rev}^{(1)} = \fr c d t_{\rm rev}^{(2)} 
= \fr e f t_{\rm rev}^{(12)}
\quad ,
\label{relcdef}
\eeq
where $c$, $d$ and $e$, $f$ are pairs of relatively
prime integers.

When Eq.\ \rf{relcdef} is obeyed,
there is a time $t = t_{\rm rev}$
at which all three second-order terms 
in the phase become integer multiples of $2 \pi$.
Near this time,
a full revival occurs:
the phase is governed once more by the first-order terms,
so the wave-packet motion and shape resemble
those of the initial wave packet.
Typically,
the full revival time $t_{\rm rev}$ is a multiple
of $t_{\rm rev}^{(1)}$, $t_{\rm rev}^{(2)}$, and 
$t_{\rm rev}^{(12)}$.
However,
it may equal one or more of the three component revival times
for particular values of the integers in \rf{relcdef}.

The appearance of fractional revivals
requires that the wave function $\Psi(t)$ in Eq.\ \rf{psiexpans}
be expressible as a sum of subsidiary wave functions.
This is possible only 
at times $t=t_{\rm frac}$
that are simultaneously irreducible rational fractions
of all three revival time scales.
We therefore write the fractional-revival time scale as 
\beq
t_{\rm frac} = \fr {p_1} {q_1} t_{\rm rev}^{(1)}
= \fr {p_2} {q_2} t_{\rm rev}^{(2)}
= \fr {p_{12}} {q_{12}} t_{\rm rev}^{(12)}
\quad .
\label{tfrac}
\eeq
Here,
the pairs of integers $(p_1,q_1)$, $(p_2,q_2)$,
and $(p_{12},q_{12})$ are relatively prime.
Note that the condition \rf{relcdef}
for the appearance of a full revival 
is automatically fulfilled 
if times $t_{\rm frac}$ can be found that satisfy Eq.\ \rf{tfrac}.
The formation of full revivals
is therefore a special case in the analysis below.

If a set of integers obeying \rf{tfrac} exists,
then the issue of the appearance of fractional
revivals at $t=t_{\rm frac}$ becomes pertinent.
To make further progress,
we introduce a doubly-periodic function 
depending only on the classical time scales:
\beq
\ps_{\rm cl} (t_1,t_2) = 
\sum_{k_1, k_2} c_{k_1 k_2}
\ph_{k_1 k_2} \exp \left[ -2 \pi i
\left(  \fr {k_1 t_1} {T_{\rm cl}^{(1)}}
+ \fr {k_2 t_2} {T_{\rm cl}^{(2)}}
\right)
\right]
\quad .
\label{psicl12}
\eeq
Here,
$t_1$ and $t_2$ are dummy variables.
The function $\ps_{\rm cl} (t_1,t_2)$ is periodic in
$t_1$ and $t_2$ with periods $T_{\rm cl}^{(1)}$ and
$T_{\rm cl}^{(2)}$,
respectively:
\beq
\ps_{\rm cl} (t_1 + T_{\rm cl}^{(1)},t_2) 
= \ps_{\rm cl} (t_1,t_2)
\quad ,
\label{uno}
\eeq
\beq
\ps_{\rm cl} (t_1,t_2 + T_{\rm cl}^{(2)}) 
= \ps_{\rm cl} (t_1,t_2)
\quad .
\label{due}
\eeq
The functions $\ps_{\rm cl} (t_1,t_2)$ have no immediate 
physical significance.
However,
at the time $t = t_1 = t_2 $
the function $\ps_{\rm cl} (t,t)$ 
is equal to the wave function $\Psi(t)$
of Eq.\ \rf{psiexpans}
when the second-order terms in the phase are omitted.
Thus,
the functions $\ps_{\rm cl} (t,t)$ 
exhibit the beating of $T_{\rm cl}^{(1)}$ and $T_{\rm cl}^{(2)}$.
Note that $\Ps(0) = \ps_{\rm cl} (0,0)$,
so for small times the behavior of $\Ps(t)$
is approximately matched by that of
$\ps_{\rm cl} (t,t)$.

The functions $\ps_{\rm cl} (t_1,t_2)$ are useful
because we can prove that,
near the times $t_{\rm frac}$ given in Eq.\ \rf{tfrac},
the wave packet $\Ps(t)$ 
can be written as a sum of subsidiary waves
$\ps_{\rm cl}$ with arguments shifted relative to $t$
by certain fractions of the corresponding periods.
The proof
involves examining the cyclic properties 
in $k_1$ and $k_2$ of the second-order terms 
in the time-dependent phase of $\Ps(t)$
and demonstrating that the functions $\ps_{\rm cl}$
with shifted arguments have identical cyclic properties.
The result means that the $\ps_{\rm cl}$ with shifted arguments
form an acceptable basis for the expansion of $\Ps(t)$
at the times $t_{\rm frac}$.

To examine the cyclic properties in $k_1$ and $k_2$
of the second-order contributions to the time-dependent
phase in $\Ps(t)$ at $t = t_{\rm frac}$,
we write these contributions as
$\exp \left( -2 \pi i \th_{k_1 k_2} \right)$,
where
\beq
\th_{k_1 k_2} = \fr {p_1} {q_1} k_1^2
+ \fr {p_2} {q_2} k_2^2 + \fr {p_{12}} {q_{12}} k_1 k_2
\quad .
\label{theta12}
\eeq
We seek the minimum periods $l_1$ and $l_2$ such that
\beq
\th_{k_1 + l_1,k_2} = \th_{k_1 k_2}
\quad , \quad\quad
\th_{k_1,k_2 + l_2} = \th_{k_1 k_2}
\quad .
\label{tre}
\eeq
Note that the relations
\beq
t_{\rm rev}^{(1)} = \fr {r_1} {s_1} t_{\rm rev}^{(12)}
\quad , \quad\quad
t_{\rm rev}^{(2)} = \fr {r_2} {s_2} t_{\rm rev}^{(12)}
\quad 
\label{quattro}
\eeq
follow from \rf{tfrac} with
\beq
\fr {r_1} {s_1} = \fr {q_1 p_{12}} {p_1 q_{12}}
\quad , \quad\quad
\fr {r_2} {s_2} = \fr {q_2 p_{12}} {p_2 q_{12}}
\quad .
\label{cinque}
\eeq
The periods $l_1$ and $l_2$ must then satisfy
\beq
\fr {p_1} {q_1} {l_1^2} + \fr {2 p_1} {q_1} l_1 k_1 
+ \fr {p_1 r_1} {q_1 s_1} l_1 k_2 = 0
\quad\quad ({\rm mod}\, 1)
\quad ,
\label{cond1}
\eeq
\beq
\fr {p_2} {q_2} {l_2^2} + \fr {2 p_2} {q_2} l_2 k_2  
+ \fr {p_2 r_2} {q_2 s_2} l_2 k_1 = 0
\quad\quad ({\rm mod}\, 1)
\quad .
\label{cond2}
\eeq
These relations can be satisfied by choosing
$l_1 = q_1 s_1$ and $l_2 = q_2 s_2$.
However,
smaller values might exist if cancellations
occur in some ratios.

The second-order contributions 
$\th_{k_1 k_2}$ are cyclic in $k_1$ and $k_2$ with
periods $l_1$ and $l_2$, respectively.
The functions $\ps_{\rm cl} (t + {s_1} T_{\rm cl}^{(1)}/{l_1},
t + {s_2} T_{\rm cl}^{(2)}/{l_2})$ with shifted arguments
have the same periodicities in $k_1$ and $k_2$.
Here,
$\ps_{\rm cl}$ is defined in Eq.\ \rf{psicl12}
and $s_1 = 0,1,\ldots,l_1-1$ and
$s_2 = 0,1,\ldots,l_2-1$.
This suggests the set of functions $\ps_{\rm cl}$
with shifted arguments
can be used as a basis for an expansion
of $\Psi(t)$ 
near the time $t_{\rm frac}$.

Explicitly,
the wave packet at the times $t \approx t_{\rm frac}$
can be written as
\beq
\Psi(t) = \sum_{s_1 = 0}^{l_1-1} \sum_{s_2 = 0}^{l_2-1}
a_{s_1 s_2} \ps_{\rm cl} (t + \fr {s_1} {l_1} T_{\rm cl}^{(1)},
t + \fr {s_2} {l_2} T_{\rm cl}^{(2)})
\quad .
\label{fracrev}
\eeq
The integers $l_1$ and $l_2$ depend on the values of
$t_{\rm frac}$.
They are obtained by solving the conditions 
\rf{cond1} and \rf{cond2}.
The coefficients $a_{s_1 s_2}$ are given by 
\beq
a_{s_1 s_2} = \fr 1 {l_1 l_2}
\sum_{\ka_1 = 0}^{l_1-1} \sum_{\ka_2 = 0}^{l_2-1}
\exp \left( - 2 \pi i \th_{\ka_1 \ka_2} \right)
\exp \left( 2 \pi i \fr {s_1} {l_1} \ka_1 \right)
\exp \left( 2 \pi i \fr {s_2} {l_2} \ka_2 \right)
\quad ,
\label{ass}
\eeq
where $\th_{\ka_1 \ka_2}$ is given by  
Eq.\ \rf{theta12}.
If the expressions for $a_{s_1 s_2}$ and 
the $\ps_{\rm cl}$ are substituted into Eq.\ \rf{fracrev},
the expansion reduces to Eq.\ \rf{psiexpans}
evaluated at the time $t=t_{\rm frac}$,
which completes the proof.

The expansion in \rf{fracrev} shows that 
the wave packet can be written as a sum of
subsidiary wave functions at $t \approx t_{\rm frac}$.
However,
it can be seen from Eq.\ \rf{fracrev}
that the subsidiary wave functions
typically have distinct shifts in the two arguments
and hence do \it not \rm follow 
the initial evolution of $\Ps(t)$,
which is given by $\ps_{\rm cl}(t,t)$.
Nonetheless,
if near these times the shifted subsidiary wave functions
are macroscopically distinct and localized in space,
then $\Ps(t)$ has a periodicity 
that is a fraction of the classical time periods.

The detailed behavior of the fractional revivals 
is governed by the form of the expansion coefficients
$a_{s_1 s_2}$ given in Eq.\ \rf{ass}.
These coefficients depend on the function
$\th_{k_1 k_2}$ defined in Eq.\ \rf{theta12},
which in turn depends on the commensurability
of the three component revival times.
An interesting special case
occurs if the cross-derivative term in 
the energy expansion vanishes,
$ {\partial^2 E}/{\partial n_1 \partial n_2} = 0$,
whereupon there is no cross-revival time $t_{\rm rev}^{(12)}$.
The coefficients $a_{s_1 s_2}$ can then be written
as a product $a_{s_1}^{(1)} a_{s_2}^{(2)}$,
where $a_{s_1}^{(1)}$ and $a_{s_2}^{(2)}$ each separately 
have the form of expansion coefficients for
the conventional fractional revivals
\cite{ap}.
Nonvanishing values of $a_{s_1}^{(1)}$ 
acquire the same norm for all $s_1$.
The same holds for $a_{s_2}^{(2)}$.
This means that for zero cross-revival time
the coefficients $a_{s_1 s_2}$ 
are equal to a product of
the corresponding coefficients for systems 
with conventional fractional revivals.
Nonetheless,
the function $\ps_{\rm cl}$ in Eq.\ \rf{psicl12}
cannot be separated into a product 
of one-dimensional functions
unless there is no entanglement 
of the states in the superposition.
Only in this special case
can the fractional revivals for systems
with energies that depend on two quantum numbers 
be written as a product of two conventional fractional revivals.
We refer to superpositions having these special properties
as totally separable wave packets.
For such packets, 
it follows that if $N_1$ subsidiary wave packets 
appear for one component while $N_2$ appear for the other,
then the fractional revival for the full system
consists of $N_1N_2$ subsidiary wave packets.

If the cross-revival time $t_{\rm rev}^{(12)}$ is nonzero,
the expansion coefficients $a_{s_1 s_2}$ cannot be
written as a product $a_{s_1}^{(1)} a_{s_2}^{(2)}$.
Their form is more complicated,
and typically they have different norms
for distinct $s_1$ and $s_2$.
The fractional revivals are therefore
less likely to cause distinctive periodicities in the
autocorrelation function.
However,
for certain circumstances only a small number 
of subsidiary wave functions enter the expansion.
This can happen,
for example,
if the values of $l_1$ and $l_2$ are small
or if some of the $a_{s_1 s_2}$ coefficients vanish.
In this case,
fractional periodicities should appear 
in the autocorrelation function.

Typically,
the conditions \rf{tfrac}
for the existence of a time $t_{\rm frac}$
at which fractional revivals can occur
are quite restrictive.
For most values of $t_{\rm rev}^{(1)}$, 
$t_{\rm rev}^{(2)}$, and $t_{\rm rev}^{(12)}$,
no solution exists.
However,
for many systems with energies that
depend on two quantum numbers
and that are of experimental interest,
there is at least one tunable parameter
such as a field strength or a length scale.
If the energy depends on the tunable parameter,
then so in general do the component revival times.
It may therefore be feasible to tune 
the system so that a particular fractional revival exists.
Since only certain commensurabilities are likely to hold,
a partial spectrum of fractional revivals should be seen.
For different values of the tunable parameter,
we expect different partial spectra of fractional revivals.

As an example of a totally separable wave packet,
consider a superposition of energy states for
a particle in a two-dimensional box 
with periodic boundary conditions.
The box is taken to have lengths $L_1$
and $L_2$ in the $x$ and $y$ directions,
respectively,
and the mass of the particle is chosen to be 
that of the electron,
i.e.,
one in atomic units.
The eigenfunctions are separable in $x$ and $y$,
with eigenenergies 
$E_{n_1 n_2} = 2\pi^2 (n_1^2 L_1^{-2} + n_2^2 L_2^{-2})$.
The wave packet is taken as
a superposition of states centered on
$\bar n_1$ and $\bar n_2$ with weighting coefficients
$c_{n_1 n_2} = c_{n_1}^{(1)} c_{n_2}^{(2)}$,
where $\vert c_{n_1}^{(1)} \vert^2$ and 
$\vert c_{n_2}^{(2)} \vert^2$ are both gaussian functions
with widths $\si_1$ and $\si_2$,
respectively.

Since the energy separates, 
there is no cross-revival time.
Furthermore,
since more than two derivatives of $E_{n_1 n_2}$ 
vanish identically,
there are no superrevivals for this system.
This means the full revivals
are perfect revivals for which
the absolute square of the autocorrelation function
is one 
\cite{bkp}.
The only nonzero revival times are
$T_{\rm cl}^{(1)} = {L_1^2}/{2 \pi \bar n_1}$,
$T_{\rm cl}^{(2)} = {L_2^2}/{2 \pi \bar n_2}$,
$t_{\rm rev}^{(1)} ={L_1^2}/{\pi}$, and
$t_{\rm rev}^{(2)} ={L_2^2}/{\pi}$.
The commensurability of the revival times may be tuned
by adjusting the ratio ${L_1^2}/{L_2^2}$.

For definiteness,
consider a superposition with
$\bar n_1 = \bar n_2 = 18 $ 
in a system with
$L_1 = \sqrt{3}/2$ and $L_2 = 1$.
The ratio of the revival times is then
$t_{\rm rev}^{(1)}/t_{\rm rev}^{(2)} = 3/4$.
We expect a full revival at $t=t_{\rm rev}
= 4 t_{\rm rev}^{(1)} = 3 t_{\rm rev}^{(2)}
= 3/\pi \simeq 0.95$ in atomic units.
Figure 1 shows the absolute square of the
autocorrelation function 
$\vert A(t) \vert^2 = \vert \vev{\Ps(0)
\vert \Ps(t)} \vert^2$
as a function of time.
A perfect full revival is evident at $t=t_{\rm rev}$.
At $t = t_{\rm rev}^{(1)} \simeq 0.24$ a.u.\
and $t = 3 t_{\rm rev}^{(1)} \simeq 0.72$ a.u.,
corresponding to the choices
${p_1}/{q_1} = 1$,
${p_2}/{q_2} = 3/4$, and 
${p_1}/{q_1} = 3$,
${p_2}/{q_2} = 9/4$,
respectively,
two subsidiary wave functions form.
At these times $\vert A(t) \vert^2 = 1/2$,
as can be seen in Fig.\ 1.
At $t = 2 t_{\rm rev}^{(1)} \simeq 0.48$ a.u.,
corresponding to 
${p_1}/{q_1} = 2$ and
${p_2}/{q_2} = 3/2$,
one wave function forms in each component.
However,
this time $t = t_{\rm rev}^{(2)}/2$,
which means the second-component subsidiary wave
is a half cycle out of phase.
Thus,
as seen in Fig.\ 1,
the autocorrelation $A(t)$ is approximately zero at
this time.

As a second explicit example,
set $L_1 = 1$ and $L_2 = \sqrt{3}$,
keeping $\bar n_1 = \bar n_2 = 18$.
These dimensions produce the ratio
$t_{\rm rev}^{(1)}/t_{\rm rev}^{(2)} = 1/3$,
and as a result different types of fractional
revivals will occur.
The autocorrelation function for this case
is shown in Fig.\ 2.
A perfect full revival occurs at 
$t=t_{\rm rev} = 3 t_{\rm rev}^{(1)} = 4 t_{\rm rev}^{(2)}
= 3/\pi \simeq 0.95$ in atomic units.
Fractional revivals consisting of three subsidiary
wave functions form at multiples of
$t_{\rm rev}^{(1)}/2 \simeq 0.16$ a.u.,
with the exception of the time $3 t_{\rm rev}^{(1)}/2$.
At this time,
${p_1}/{q_1} = 3/2$ and
${p_2}/{q_2} = 1/2$,
and each component consists of a single wave packet
forming one-half cycle out of phase with the classical motion.
This does \it not \rm result in a perfect revival because the
cycles are not equal,
$T_{\rm cl}^{(1)}/2 \ne T_{\rm cl}^{(2)}/2$,
so the two components are out of phase.
Additional fractional revivals consisting of four subsidiary
wave functions are visible at times
$t = 3 t_{\rm rev}^{(1)}/4 \simeq 0.24$ a.u.\ and 
$t = 9 t_{\rm rev}^{(1)}/4 \simeq 0.72$ a.u.

The analysis in this paper is experimentally testable.
As an explicit example of a physical system 
for which the energy depends on two quantum numbers 
and a cross-revival time exists,
we can consider a Stark wave packet.
This type of packet can be produced 
using a static electric field applied to an atom,
which shifts and splits the energy levels.
Applying a short laser pulse 
then creates a coherent superposition of Stark levels.
The energies in atomic units 
for hydrogen and a weak electric field
are $E_{n k} = - 1/{2 n^2} + 3nkF/2$,
where $n$ is the principal quantum number,
$k$ is the difference $n_1 - n_2$ 
between the parabolic quantum numbers,
and $F$ is the magnitude of the electric-field strength.

Stark wave packets have unique features because the  
quantum number $k$ is
even or odd according to whether $n$ is odd or even
and because for fixed $n$ 
the adjacent $k$ values differ by two units.
These properties introduce some additional complications in the
analysis of fractional revivals and 
produce novel interference patterns in the interferograms
of Stark wave packets
\cite{swp}.
A second-order Taylor expansion of the energy $E_{nk}$ 
with respect to $n$ and $k$ can still be performed,
but some changes occur in the definitions of the 
controlling time scales.
Since the quantum number $k$ is either even or odd,
a Stark wave packet effectively evolves as two separate wave packets.
At the fractional revival times,
the packet separates into sums 
over sets of subsidiary wave functions 
consisting of superpositions of even-$n$ and odd-$n$ states.
These wave functions evolve with distinct relative phases
and obey distinct periodicity relations.
This results in additional interference patterns
in the autocorrelation function,
which can be tested experimentally.

\vglue 0.6cm
One of us (R.B.) would like to thank Colby College
for a Science Division grant.
This work is supported in part by the National
Science Foundation under grant number PHY-9503756.

\vglue 0.6cm
{\bf\noindent REFERENCES}
\vglue 0.4cm

\vfill\eject

\baselineskip=16pt
{\bf\noindent FIGURE CAPTIONS}
\vglue 0.4cm

\begin{description}
 
\item[{\rm Fig.\ 1:}]
The absolute square of the autocorrelation function
is plotted as a function of time in atomic units
for a particle in a two-dimensional box with
$\bar n_1 = \bar n_2 = 18$,
$L_1 = \sqrt{3}/2$, and $L_2 = 1$.

\item[{\rm Fig.\ 2:}]
The absolute square of the autocorrelation function
is plotted as a function of time in atomic units
for a particle in a two-dimensional box with
$\bar n_1 = \bar n_2 = 18$,
$L_1 = 1$, and $L_2 = \sqrt{3}$.

\end{description}

\vfill
\eject


\begin{thebibliography}{xx}

\bibitem{yeazell3}
J.A. Yeazell and C.R. Stroud,
Phys. Rev. A {\bf 43}, 5153 (1991).

\bibitem{meacher}
D.R. Meacher, P.E. Meyler, I.G. Hughes, and P. Ewart,
J. Phys. B {\bf 24}, L63 (1991).

\bibitem{wals}
J. Wals, H.H. Fielding, J.F. Christian, L.C. Snoek,
W.J. van der Zande,
and H.B. van Linden van den Heuvell,
Phys. Rev. Lett. {\bf 72}, 3783 (1994).

\bibitem{stolow}
M.J.J. Vrakking, D.M. Villeneuve, and A. Stolow,
Phys. Rev. A {\bf 54}, R37 (1996).

\bibitem{az}
G. Alber, H. Ritsch, and P. Zoller,
Phys. Rev. A {\bf 34}, 1058 (1986);
G. Alber and P. Zoller,
Phys. Rep. {\bf 199}, 231 (1991).

\bibitem{ap}
I.Sh. Averbukh and N.F. Perelman,
Phys. Lett. {\bf 139A}, 449 (1989).

\bibitem{bkp}
R. Bluhm, V.A. Kosteleck\'y, and J. Porter,
Am.\ J.\ Phys.\ {\bf 64}, 944 (1996) (quant-ph/9510029).

\bibitem{sr}
R. Bluhm and V.A. Kosteleck\'y,
Phys.~Rev.~A {\bf 50}, R4445 (1994) (hep-ph/9410325);
Phys.~Lett.~A {\bf 200}, 308 (1995) (quant-ph/9508024);
Phys.~Rev.~A {\bf 51}, 4767 (1995) (quant-ph/9506009).

\bibitem{ps}
J. Parker and C.R. Stroud,
Phys. Rev. Lett. {\bf 56}, 716 (1986).

\bibitem{peres}
A. Peres,
Phys.~Rev.~A {\bf 47}, 5196 (1993).

\bibitem{sqdt}
V.A. Kosteleck\'y and M.M. Nieto, 
Phys. Rev. Lett. {\bf 53}, 2285 (1984);
Phys. Rev. A {\bf 32}, 1293, 3243 (1985).

\bibitem{squeezed}
R. Bluhm and V.A. Kosteleck\'y,
Phys. Rev. A {\bf 48}, R4047 (1993) (quant-ph/9508019);
Phys. Rev. A {\bf 49}, 4628 (1994) (quant-ph/9508020);
R. Bluhm, V.A. Kosteleck\'y, and B. Tudose,
Phys. Rev. A {\bf 52}, 2234 (1995) (quant-ph/9509010);
Phys. Rev. A {\bf 53}, 937 (1996) (quant-ph/9510023).

\bibitem{noordam}
A. ten Wolde, L.D. Noordam, A. Lagendijk, 
and H.B. van Linden van den Heuvell,
Phys.~Rev.~A {\bf 40}, 485 (1989);
L.D. Noordam, A. ten Wolde, A. Lagendijk, 
and H.B. van Linden van den Heuvell,
Phys.~Rev.~A {\bf 40}, 6999 (1989).

\bibitem{walther}
J.A. Yeazell, G. Raithel, L. Marmet, H. Held, and H. Walther,
Phys. Rev. Lett. {\bf 70}, 2884 (1993).

\bibitem{broers}
B. Broers, J.F. Christian, J.H. Hoogenraad, W.J. van der Zande,
H.B. van Linden van den Heuvell, and L.D. Noordam,
Phys. Rev. Lett. {\bf 71}, 344 (1993).

\bibitem{lankhui}
G.M. Lankhuijzen and L.D. Noordam, 
Phys.~Rev.~A {\bf 52}, 2016 (1995).

\bibitem{bucksbaum}
M.L. Naudeau, C.I. Sukenik, and P.H. Bucksbaum,
``Core Scattering of Stark Wavepackets,''
University of Michigan preprint. 

\bibitem{swp}
R. Bluhm, V.A. Kosteleck\'y, and B. Tudose,
Phys. Rev. A, in press (quant-ph/9608040).

\end{thebibliography}
\end{document}